\documentclass[aps,pra,superscriptaddress,twocolumn]{revtex4}
\usepackage[utf8x]{inputenc}
\usepackage{float}
\usepackage{bm}
\usepackage{graphicx,amsmath,latexsym,amssymb}
\usepackage{color}
\usepackage{dcolumn}
\begin{document}
\title{
Nonconservative dipole forces in a two-atom system with one atom excited}
\author{J. S\'anchez-C\'anovas} 
\affiliation{Departamento de F\'isica Te\'orica, At\'omica y \'Optica and IMUVA,  Universidad de Valladolid, Paseo Bel\'en 7, 47011 Valladolid, Spain}
\author{M. Donaire} 
\email{manuel.donaire@uva.es}
\affiliation{Departamento de F\'isica Te\'orica, At\'omica y \'Optica and IMUVA,  Universidad de Valladolid, Paseo Bel\'en 7, 47011 Valladolid, Spain}

\begin{abstract}
We compute the nonconservative electric dipole forces between two atoms, one of which is initially excited. These forces 
derive from the time variation of the longitudinal electromagnetic momentum.  
In contrast to the conservative van-der-Waals forces, the nonconservative forces posses components orthogonal to the interatomic axis. Thus, despite being 
several orders of magnitude smaller than van-der-Waals', they might be accessible experimentally for the case of two identical atoms. 
In addition, as with the van-der-Waals forces, the nonconservative forces present nonreciprocal constituents which result 
in a net force on the two-atom system. 
We offer an estimate of the spatial displacement caused by the nonconservative forces on a binary system of Hydrogen atoms.
\end{abstract}
\maketitle

\section{Introduction}  
 Dispersion forces between neutral atoms in the electric dipole approximation are generally referred to as van-der-Waals (vdW) forces 
 \cite{vdW,WileySipe,Milonnibook,Miltonbook,BuhmannbookI,BuhmannbookII,Craigbook,BuhmannScheel}. They are the result of the coupling of the quantum fluctuations 
 of the electromagnetic (EM) 
field in its vacuum state with the dipole fluctuations of the atomic charges in stable or metastable states. For a system of atoms in its ground state the vdW 
forces can be computed applying the usual techniques of stationary quantum perturbation theory. Those forces are conservative and reciprocal, 
and can be expressed in terms of the atomic polarizabilities \cite{Milonnibook,Miltonbook,BuhmannbookI,BuhmannbookII,Craigbook,BuhmannScheel}. 
 In contrast, when atoms are excited, it has been proved in Refs.\cite{Passante,Berman,MePRL,Milonni,Pablo,MePRAvdW,Sherkunov1,Sherkunov2} that a fully stationary treatment 
 is insufficient to account for the incoherent dynamics of a metastable system. In particular, for a system of two dissimilar atoms with one of them initially 
 excited, the time-dependent approaches of Refs.\cite{Berman,Pablo,MePRAvdW}, in the adiabatic limit, have shown that  
while the resonant component of the vdW force upon 
the excited atom oscillates in space, the resonant force on the de-excited atom decreases monotoncially with the interatomic distance. The nonreciprocity 
of the vdW forces results in an apparent violation of the classical action-reaction principle and the conservation of total momentum, which would be in contradiction 
with the invariance of the system under global spatial translation. This apparent contradiction was solved in Ref.\cite{My_Net_PRA}, where it was shown that the 
\emph{missing} momentum was carried by the photons which mediate the interaction which, ultimately, causes the directioality of spontaneous emission 
when the system gets de-excited. The vdW forces between dissimilar atoms, either resonant or off-resonant, are all quasi-stationary for an adiabatic excitation, 
and can be expressed in terms of the gradients of the expectation values of the interaction potentials \cite{MePRAvdW,My_Net_PRA}, hence reflecting their conservative 
nature. 

As for the case of a binary system of identical atoms, with one of them initially excited,  
it has been shown in Ref.\cite{Julio_PRA1} that the vdW forces are inherently time-dependent and grow linearly with time in the perturbative regime. This 
is so because, on the one hand, the system is degenerate and an adiabatic excitation is not feasible. In fact, a sudden excitation is generally a good approximation 
to model the preparation of the initial state of the system. On the other hand, that excited state is highly 
nonstationary since its dynamics comprises both the coherent transfer of the excitation between the atoms and its incoherent decay through spontaneous emission. 


Besides, it was found in Ref.\cite{Julio_PRA1} that, still in the electric dipole approximation, in addition to the conservative vdW forces  
 nonconservative forces may arise from the time variation of the EM longitudinal momentum \cite{Cohen,Cohen_Kawkathesis_KawkavanTiggelen_Dippel,
KawkavanTiggelenRikken,Kawkathesis,Baxter1,Baxter2}. Since the  
interaction between two atoms becomes time-dependent when one of them is excited nonadiabatically, so does the EM longitudinal momentum of the system. Thus, 
 nonconservative forces arise in a binary system if excited nonadiabatically and, as for the case of the conservative vdW forces, their strength is greater for the case of 
 identical atoms. 
 

In this article we aim at computating the nonconservative forces on a binary system of identical two-level atoms.  We will show that they contain components 
orthogonal to the interatomic axis which might be accessible experimentally. We will estimate the spatial displacement caused by the nonconservative 
dipole forces on a binary system of Hydrogen atoms. In addition, we will show that such forces 
posses nonreciprocal terms which result in a net nonconservative force upon the two-atom system.

\section{Fundamentals of the approach}\label{fundam}
Let us take a system of two two-level atoms, $A$ and $B$, located a distance $R$ apart. In the first place, let us consider dissimilar atoms with resonance 
frequencies $\omega_{A}$ and $\omega_{B}$, detuning $\Delta_{AB}=\omega_{A}-\omega_{B}$, natural linewidths $\Gamma_{A}$ and $\Gamma_{B}$, with the excited level 
of atom $B$ being n-fold degenerate. Since we are ultimately interested in the identical atoms limit, 
$|\Delta_{AB}|\ll\Gamma_{A}$, $\Gamma_{A}\rightarrow\Gamma_{B}$, atom $A$ is assumed to be suddenly excited with an external 
field of strength $\Omega\gg|\Delta_{AB}|$. Thus, the state of the system at time 0 is $|\Psi(0)\rangle=|A_{+}\rangle\otimes|B_{-}\rangle\otimes|0_{\gamma}\rangle$, 
where $|A_{+}\rangle$ is the excited state of atom $A$, $|(A,B)_{-}\rangle$ denote the ground states of the atoms $A$ and $B$, respectively, 
$|0_{\gamma}\rangle$ is the EM vacuum state, and 
the states of the n-fold degenerate excited state of atom $B$ will be denoted by $\{|b\rangle\}$.
At time $T>0$ the state of the two-atom-EM field system writes $|\Psi(T)\rangle=\mathbb{U}(T)|\Psi(0)\rangle$, where $\mathbb{U}(T)$ denotes the time propagator in 
the Schr\"odinger representation,
\begin{align}
\mathbb{U}(T)&=\textrm{T-exp}\Bigl\{-i\hbar^{-1}\int_{0}^{T}\textrm{d}t\:H\Bigr\},\label{UT}\\
H&=\mathcal{T}+H_{A}+H_{B}+H_{EM}+W.\nonumber
\end{align}
In this equation $\mathcal{T}=m_{A}|\dot{\mathbf{R}}_{A}|^{2}/2+m_{B}|\dot{\mathbf{R}}_{B}|^{2}/2$ is the kinetic energy of the center of mass of the atomic system, 
with $m_{A,B}$ being the atomic masses and $\mathbf{R}_{A,B}$ the position vectors of the centers of mass of each atom. $H_{A}+H_{B}$ is the free Hamiltonian of the 
internal atomic states, $\hbar\omega_{A}|A_{+}\rangle\langle A_{+}|+\sum_{b}\hbar\omega_{B}|b\rangle\langle b|$, while the Hamiltonian of the free EM field is 
$H_{EM}=\sum_{\mathbf{k},\boldsymbol{\epsilon}}\hbar\omega(a^{\dagger}_{\mathbf{k},\boldsymbol{\epsilon}}a_{\mathbf{k},\boldsymbol{\epsilon}}+1/2)$,
where $\omega=ck$ is the photon frequency, and the operators $a^{\dagger}_{\mathbf{k},\boldsymbol{\epsilon}}$ and $a_{\mathbf{k},\boldsymbol{\epsilon}}$ are the 
creation and annihilation operators of photons with momentum $\hbar\mathbf{k}$ and polarization $\boldsymbol{\epsilon}$, respectively. Finally, the interaction 
Hamiltonian in the electric dipole approximation reads $W=W_{A}+W_{B}$, with 
\begin{eqnarray}
W_{A,B}&=&-\mathbf{d}_{A,B}\cdot\mathbf{E}(\mathbf{R}_{A,B})+\Bigl[\mathbf{P}_{A,B}\cdot[\mathbf{d}_{A,B}\times\mathbf{B}(\mathbf{R}_{A,B})]\nonumber\\
&+&[\mathbf{d}_{A,B}\times\mathbf{B}(\mathbf{R}_{A,B})]\cdot\mathbf{P}_{A,B}\Bigr]/2m_{A,B}.\label{WAB}
\end{eqnarray}
The first term on the right hand side of this equation is the usual EM interaction of an electric dipole with the electric field, whereas the second term is the 
so-called R\"ontgen's 
 term, which accounts for the coupling of the canonical conjugate momentum of each atom, $\mathbf{P}_{A,B}$, to the electromagnetic vector 
 potential, $\mathbf{A}$, in the electric dipole approximation \cite{Kawkathesis,Baxter1,Baxter2}. The electric dipole operators 
 are denoted by $\mathbf{d}_{A,B}$, and $\mathbf{E}(\mathbf{R}_{A,B})$, $\mathbf{B}(\mathbf{R}_{A,B})$ are the quantum 
 electric and magnetic field operators in Schr\"odinger's representation, respectively. In terms of the EM vector potential,
 the electric and magnetic fields, $\mathbf{E}(\mathbf{R}_{A,B})=-\partial_{t}\mathbf{A}(\mathbf{R}_{A,B},t)|_{t=0}$, 
 $\mathbf{B}(\mathbf{R}_{A,B})=\boldsymbol{\nabla}_{A,B}\times\mathbf{A}(\mathbf{R}_{A,B})|_{t=0}$, can be written as sums over normal modes 
 \cite{Milonnibook,Craigbook},
\begin{align}\label{AQ}
\mathbf{E}&(\mathbf{R}_{A,B})=\sum_{\mathbf{k}}\mathbf{E}^{(-)}_{\mathbf{k}}(\mathbf{R}_{A,B})+\mathbf{E}^{(+)}_{\mathbf{k}}(\mathbf{R}_{A,B})\nonumber\\
&=i\sum_{\mathbf{k},\boldsymbol{\epsilon}}\sqrt{\frac{\hbar ck}{2\mathcal{V}\epsilon_{0}}}
[\boldsymbol{\epsilon}a_{\mathbf{k},\boldsymbol{\epsilon}}e^{i\mathbf{k}\cdot\mathbf{R}_{A,B}}-\boldsymbol{\epsilon}^{*}a^{\dagger}_{\mathbf{k},\boldsymbol{\epsilon}}e^{-i\mathbf{k}\cdot\mathbf{R}_{A,B}}],\nonumber\\
\mathbf{B}&(\mathbf{R}_{A,B})=\sum_{\mathbf{k}}\mathbf{B}^{(-)}_{\mathbf{k}}(\mathbf{R}_{A,B})+\mathbf{B}^{(+)}_{\mathbf{k}}(\mathbf{R}_{A,B})\nonumber\\
&=i\sum_{\mathbf{k},\boldsymbol{\epsilon}}\sqrt{\frac{\hbar}{2ck\mathcal{V}\epsilon_{0}}}\mathbf{k}\times
[\boldsymbol{\epsilon}a_{\mathbf{k},\boldsymbol{\epsilon}}e^{i\mathbf{k}\cdot\mathbf{R}_{A,B}}-\boldsymbol{\epsilon}^{*}a^{\dagger}_{\mathbf{k},\boldsymbol{\epsilon}}
e^{-i\mathbf{k}\cdot\mathbf{R}_{A,B}}],\nonumber
\end{align}
where $\mathcal{V}$ is a generic volume and $\mathbf{E}^{(\mp)}_{\mathbf{k}}$, $\mathbf{B}^{(\mp)}_{\mathbf{k}}$ denote the annihilation/creation electric and 
magnetic field operators of photons of momentum $\hbar\mathbf{k}$, respectively. Hereafter we will drop R\"ontgen's term from Eq.(\ref{WAB}), for its contributions 
to Eq.(\ref{UT}) as well as to any time-dependent expectation value are of orders $\dot{R}_{A,B}/c$ and $\mathbf{d}_{A,B}\cdot\mathbf{E}(\mathbf{R}_{A,B})/m_{A,B}$ 
smaller than the contributions of the first term in Eq.(\ref{WAB}) \cite{MeQFriction}.

Next, considering $W$ as a perturbation to the free Hamiltonians, the unperturbed time propagator for atom and free photon states is 
$\mathbb{U}_{0}(t)=\exp{[-i\hbar^{-1}(\mathcal{T}+H_{A}+H_{B}+H_{EM})t]}$. In terms of $W$ and $\mathbb{U}_{0}$, $\mathbb{U}(T)$ admits an expansion in powers of  $W$ which 
can be developed out of its time-ordered exponential expression,
\begin{equation}\label{U}
\mathbb{U}(T)=\mathbb{U}_{0}(T)\:\textrm{T-exp}\int_{0}^{T}(-i/\hbar)\mathbb{U}_{0}^{\dagger}(t)\:W\:\mathbb{U}_{0}(t)\textrm{d}t,
\end{equation}
which can be written as $\mathbb{U}(T)=\mathbb{U}_{0}(T)+\sum_{n=1}^{\infty}\delta\mathbb{U}^{(n)}(T)$, with $\delta\mathbb{U}^{(n)}$ being the term of order $W^{n}$. 

The system posseses a conserved total momentum, $\mathbf{K}=\mathbf{P}_{A}+\mathbf{P}_{B}+\mathbf{P}_{\perp}^{\gamma}$, with 
$\mathbf{P}_{\perp}^{\gamma}=\sum_{\mathbf{k},\mathbf{\epsilon}}\hbar\mathbf{k}\:a^{\dagger}_{\mathbf{k},\mathbf{\epsilon}}a_{\mathbf{k},\mathbf{\epsilon}}$ 
being the transverse EM momentum, which satisfies 
$[H,\mathbf{K}]=\mathbf{0}$  \cite{Cohen,Cohen_Kawkathesis_KawkavanTiggelen_Dippel,Kawkathesis}. 
Further, if the charges  $\{q_{i}\}$ within the atoms are considered individually at positions $\{\mathbf{r}_{i}\}$, the total canonical 
conjugate momentum can be written as 
\begin{equation}
\mathbf{P}_{A}+\mathbf{P}_{B}=m_{A}\dot{\mathbf{R}}_{A}+m_{B}\dot{\mathbf{R}}_{B}+\sum_{i}q_{i}\mathbf{A}(\mathbf{r}_{i}),\label{conjugate}
\end{equation}
where the first two terms are the kinetic momenta of the centers of mass of each atom, and the momentum within the summation symbol is referred to as longitudinal 
EM momentum \cite{Cohen,Cohen_Kawkathesis_KawkavanTiggelen_Dippel}, 
$\mathbf{P}_{\parallel}^{\gamma}=\sum_{i}q_{i}\mathbf{A}(\mathbf{r}_{i})$. This is the EM momentum which results from the combination 
of the Coulomb electric field and the magnetic field generated by the internal motion of the atomic charges \cite{Cohen,JPCM}. Lastly, in the electric dipole approximation, $\mathbf{P}_{\parallel}^{\gamma}$ reads 
 $\mathbf{P}_{\parallel}^{\gamma}\simeq-\mathbf{d}_{A}\times\mathbf{B}(\mathbf{R}_{A})-\mathbf{d}_{B}\times\mathbf{B}(\mathbf{R}_{B})$ \cite{Baxter1,Baxter2}.

Following Refs.\cite{MeQFriction,My_Net_PRA}, the electric dipole force on each atom is computed applying the time derivative to the expectation value of the 
kinetic momenta of the centers of mass of each atom. Writing the latter in terms of the canonical conjugate momenta and the longitudinal EM momentum, 
in the electric dipole approximation, we arrive at
\begin{align}
\langle\mathbf{F}_{A,B}\rangle_{T}&=\partial_{T}\langle m_{A,B}\dot{\mathbf{R}}_{A,B}\rangle_{T}\label{Force}\\
&=-i\hbar\partial_{T}\langle\Psi(0)|\mathbb{U}^{\dagger}(T)\boldsymbol{\nabla}_{A,B}\mathbb{U}(T)|\Psi(0)\rangle\nonumber\\
&+\partial_{T}\langle\Psi(0)|\mathbb{U}^{\dagger}(T)\mathbf{d}_{A,B}\times\mathbf{B}(\mathbf{R}_{A,B})\mathbb{U}(T)|\Psi(0)\rangle\nonumber\\
&=-\langle \boldsymbol{\nabla}_{A,B}W_{A,B}\rangle_{T}+
\partial_{T}\langle\mathbf{d}_{A,B}\times\mathbf{B}(\mathbf{R}_{A,B})\rangle_{T}.\nonumber
\end{align}
The first term on the right hand side of the last equality corresponds to the conservative vdW forces along the interatomic axis, already computed in 
Ref.\cite{Julio_PRA1}. The second term corresponds to the nonconservative forces we are interested in, 
$\langle\mathbf{F}_{A,B}^{nc}\rangle_{T}=\partial_{T}\langle\mathbf{d}_{A,B}\times\mathbf{B}(\mathbf{R}_{A,B})\rangle_{T}$, which equal the 
time derivatives of the components of the longitudinal EM momentum, with opposite signs.

\section{Computation of the nonconservative forces} 
\begin{figure}[H]
\includegraphics[height=7.2cm,width=8.9cm,clip]{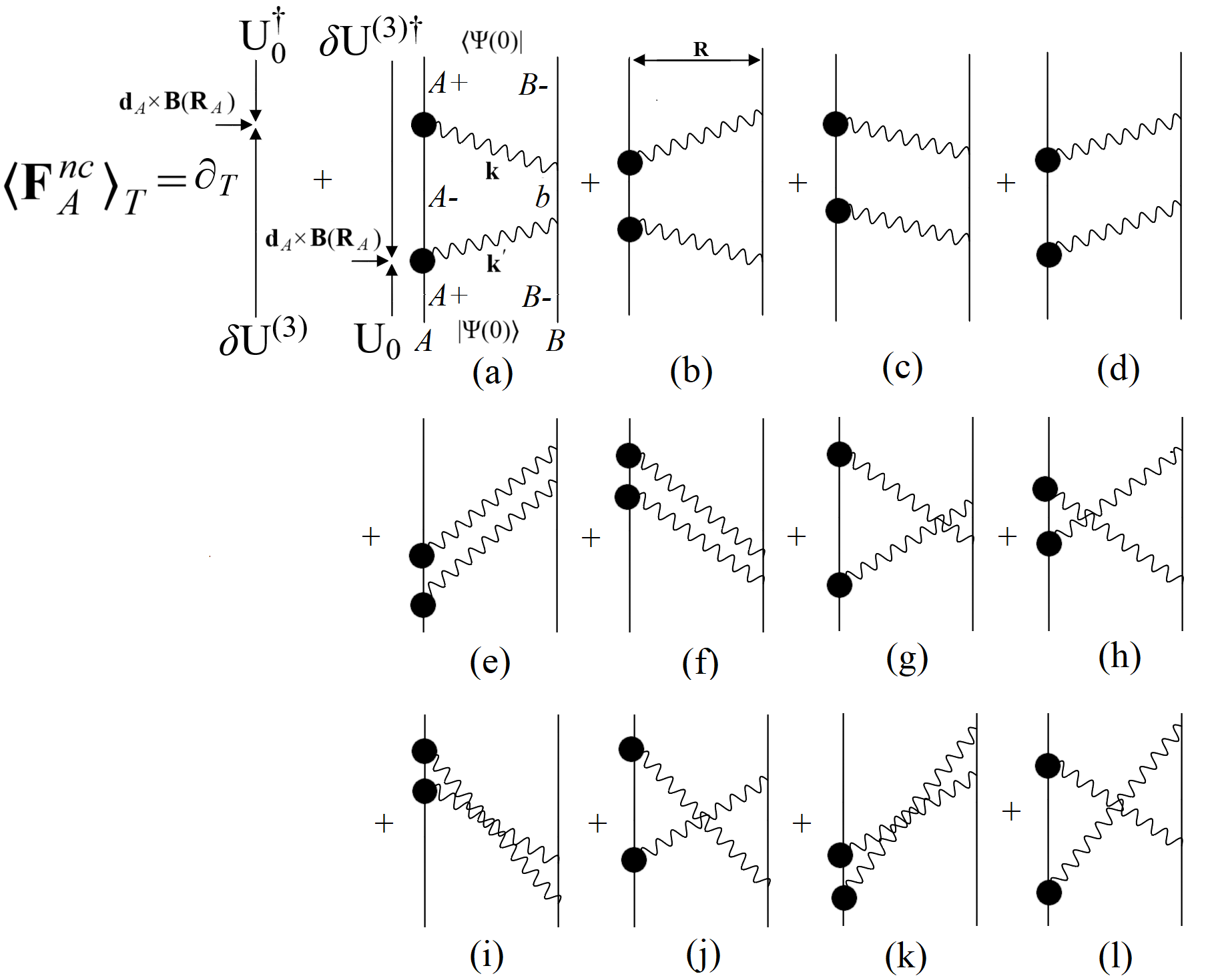}
\caption{Diagrammatic representation of the twelve  processes which contribute to $\langle\mathbf{F}_{A}^{nc}\rangle_{T}$. 
Thick straight lines stand for propagators of atomic states, while wavy lines stand for photon propagators. In diagram (a), atomic and photon states are indicated 
explicitly. The atoms $A$ and $B$ are separated by a distance $R$ along the horizontal direction, whereas time runs along the vertical. 
The big circles in black on the left of each diagram stand for the insertion of the Schr\"odinger operator $\mathbf{d}_{A}\times\mathbf{B}(\mathbf{R}_{A})$ 
whose expectation value is computed.  
Each diagram contributes with two terms, one from each of the operators inserted. 
They are sandwiched between two time propagators, $\mathbb{U}(T)$ and $\mathbb{U}^{\dagger}(T)$ (depicted by vertical arrows), which evolve the initial state 
$|\Psi(0)\rangle$ towards the observation time at which $\mathbf{d}_{A}\times\mathbf{B}(\mathbf{R}_{A})$ applies.}\label{figure1A}
\end{figure}

In what follows we compute the nonconservative forces, in the identical atoms limit, including up to two-photon exchange processes. These are depicted 
diagrammatically for $\langle\mathbf{F}_{A}^{nc}\rangle_{T}$ in Fig.\ref{figure1A}. Analogous diagrams contribute to $\langle\mathbf{F}_{B}^{nc}\rangle_{T}$. 
The contributions of each process to $\langle\mathbf{F}_{A}^{nc}\rangle_{T}$ and $\langle\mathbf{F}_{B}^{nc}\rangle_{T}$ 
are analogous to those of the vdW forces computed in Ref.\cite{Julio_PRA1}, but for the replacement of the operators $-\boldsymbol{\nabla}_{A,B}W_{A,B}$ with 
$\partial_{T}\mathbf{d}_{A,B}\times\mathbf{B}(\mathbf{R}_{A,B})$ acting upon one of the exchanged photons.
Their complete expressions are compiled in the Appendix \ref{appendix}, where the origin of each contribution  
is explained in terms of the diagrams in Fig.\ref{figure1A}. In particular, the dominant terms arise from diagram (a). Further, considering the 
identical atoms limit, $\Gamma_{A}\rightarrow\Gamma_{B}\equiv\Gamma_{0}$, $\omega_{A}\rightarrow\omega_{B}\equiv\omega_{0}$, $\Delta_{AB}/\Gamma_{0}\rightarrow0$, 
and defining the 
transition dipole moments as $\boldsymbol{\mu}_{A}=\langle A_{-}|\mathbf{d}_{A}|A_{+}\rangle$, $\boldsymbol{\mu}_{b}=\langle B_{-}|\mathbf{d}_{B}|b\rangle$, 
the leading order terms of the nonconservative forces on each atom read,
\begin{align}
&\langle  \mathbf{F}_A^{nc}\rangle_{T}=\frac{2 \omega_0^4 (1-\Gamma_0 T)e^{-\Gamma_0T}}{-c^5 \epsilon_0 \hbar} \sum_{b} \left[\mu_A^{\parallel} 
\boldsymbol{\mu}_b^{\perp}-\mu_A^{\perp}\mu_b^{\perp} \hat{\mathbf{R}}\right]\nonumber\\ 
&\times \boldsymbol{\mu}_A \cdot \left[\textrm{Re}\mathcal{G}(k_0R)\textrm{Im}\mathbb{G}(k_0R)+\textrm{Im}\mathcal{G}(k_0R)\textrm{Re}\mathbb{G}(k_0R)\right] 
\cdot \boldsymbol{\mu}_b,\nonumber\\
&\langle  \mathbf{F}_B^{nc}\rangle_{T}=\frac{2 \omega_0^4 (1-\Gamma_0 T)e^{-\Gamma_0T}}{-c^5 \epsilon_0 \hbar} \sum_{b} 
\left[\mu_b^{\parallel} \boldsymbol{\mu}_A^{\perp}-\mu_A^{\perp}\mu_b^{\perp} \hat{\mathbf{R}}\right]\nonumber\\ 
&\times \boldsymbol{\mu}_A
\cdot \left[\textrm{Re}\mathcal{G}(k_0R)\textrm{Im}\mathbb{G}(k_0R)-\textrm{Im}\mathcal{G}(k_0R)\textrm{Re}\mathbb{G}(k_0R)\right] \cdot \boldsymbol{\mu}_b,
\label{Fnc}
\end{align}
where $\mathbf{R}=\mathbf{R}_{A}-\mathbf{R}_{B}$, $\hat{\mathbf{R}}=\mathbf{R}/R$;  
$\mathbb{G}(kr)$ is the dyadic Green's function of the electric field induced at $\mathbf{r}$ by an electric dipole of frequency $ck$ at the origin, 
\begin{equation}
\mathbb{G}(kr)=\frac{k\:e^{ikr}}{-4\pi}[ \alpha/kr+i \beta/(kr)^{2}- \beta/(kr)^{3}],\label{Gee}
\end{equation}
with $ \alpha=\mathbb{I}-\hat{\mathbf{r}}\hat{\mathbf{r}}$,  $ \beta=\mathbb{I}-3\hat{\mathbf{r}}\hat{\mathbf{r}}$; 
$\mu_{A,b}^{\parallel}=\boldsymbol{\mu}_{A,b}\cdot\hat{\mathbf{R}}$, $\boldsymbol{\mu}_{A,b}^{\perp}=\alpha\cdot\boldsymbol{\mu}_{A,b}$; and 
$\mathcal{G}(kr)\mathcal{E}\cdot\hat{\mathbf{r}}$ is the 
dyadic Green's tensor of the magnetic field induced at $\mathbf{r}$ by an electric dipole of frequency $ck$ at the origin (cf. Ref.\cite{MeQFriction} and Appendix \ref{appendix}), 
with $\mathcal{E}$ being the 3-dimensional Levi-Civita tensor and
\begin{equation}
\mathcal{G}(kr)=-\frac{k e^{ikr}}{4 \pi  } \left(\frac{1}{kr}+\frac{i}{(kr)^2}\right).\label{Gem}
\end{equation}
Substituting Eqs.(\ref{Gee}) and (\ref{Gem}) into Eq.(\ref{Fnc}) one obtains that, as with the conservative vdW forces, $\langle\mathbf{F}_A^{nc}\rangle_{T}$ 
oscillates in space with wavelength $\pi/k_{0}$, 
\begin{align}
&\langle \mathbf{F}^{nc}_{A}\rangle_T=\frac{k_0^6 (1- \Gamma_0 T)e^{-\Gamma_0 T}}{-8 \pi^2 \epsilon_0  \hbar c} \sum_b 
\left[\mu_A^{\parallel}\boldsymbol{\mu}_b^{\perp}-\mu_A^{\perp}\mu_b^{\perp} \hat{\mathbf{R}}\right]\nonumber\\ 
&\times \boldsymbol{\mu}_A \cdot \Bigl[ \frac{\alpha}{(k_0R)^2}\left(\sin (2k_0R)+\frac{\cos(2k_0R)}{ k_0R}\right)+ \frac{\beta}{(k_0R)^3}\nonumber\\
&\times\left(\cos (2k_0R)-\frac{2\sin(2k_0R)}{k_0R}-\frac{\cos(2k_0R)}{(k_0R)^2}\right)\Bigr] \cdot \boldsymbol{\mu}_B, 
\end{align}
whereas $\langle\mathbf{F}_B^{nc}\rangle_{T}$ decreases monotonically with $R$,
\begin{align}
\langle \mathbf{F}^{nc}_B\rangle_T&=\frac{k_0^6 (1- \Gamma_0 T)e^{-\Gamma_0 T}}{-8 \pi^2 \epsilon_0  \hbar c} \sum_b 
\left[\mu_b^{\parallel}\boldsymbol{\mu}_A^{\perp}-\mu_A^{\perp}\mu_b^{\perp} \hat{\mathbf{R}}\right]\nonumber\\
&\times\boldsymbol{\mu}_A \cdot \left[ \frac{\beta-\alpha}{(k_0R)^3}+ \frac{\beta}{(k_0R)^5}\right] \cdot \boldsymbol{\mu}_B. 
\end{align} 

In contrast to the conservative vdW forces, the most remarkable feature of the nonconservative forces is the presence of components which are perpendicular to 
the interatomic axis $(\perp)$. Also, as with  
the vdW forces, nonconservative forces posses reciprocal and non-reciprocal components \cite{MePRAvdW,My_Net_PRA,Julio_PRA1}. 
The former, $\pm\langle\mathbf{F}_{A}^{nc}-\mathbf{F}_{B}^{nc}\rangle_{T}/2$, satisfy the ordinary 
action reaction principle; while the latter amount to a net force on the two atom system, $\langle\mathbf{F}_{A}^{nc}+\mathbf{F}_{B}^{nc}\rangle_{T}$,
\begin{align}
&\langle \mathbf{F}_A^{nc} + \mathbf{F}_B^{nc}\rangle_{T}=\frac{2\omega_0^4 (1-\Gamma_0 T)}{-c^5 \epsilon_0 \hbar e^{\Gamma_0T}}\sum_{b}
\left[\mu_A^{\parallel} \boldsymbol{\mu}_b^{\perp}+\mu_b^{\parallel} \boldsymbol{\mu}_A^{\perp} \right.\nonumber\\
&\left.-2\mu_A^{\perp}\mu_b^{\perp} \hat{\mathbf{R}}\right]\boldsymbol{\mu}_A \cdot \textrm{Re}\mathcal{G}(k_0R)\textrm{Im}\mathbb{G}(k_0R)\cdot\boldsymbol{\mu}_b
\nonumber\\
&+\left[\mu_A^{\parallel} \boldsymbol{\mu}_b^{\perp}-\mu_b^{\parallel} \boldsymbol{\mu}_A^{\perp}\right]
\boldsymbol{\mu}_A \cdot \textrm{Im}\mathcal{G}(k_0R)\textrm{Re}\mathbb{G}(k_0R)\cdot \boldsymbol{\mu}_b,\label{Fnc_net}
\end{align}
which oscillates in space as $\sim\sin{(2k_{0}R)}/(k_{0}R)^{2}$ in the retarded regime.
\begin{figure}[H]
\includegraphics[width=8.3cm,clip]{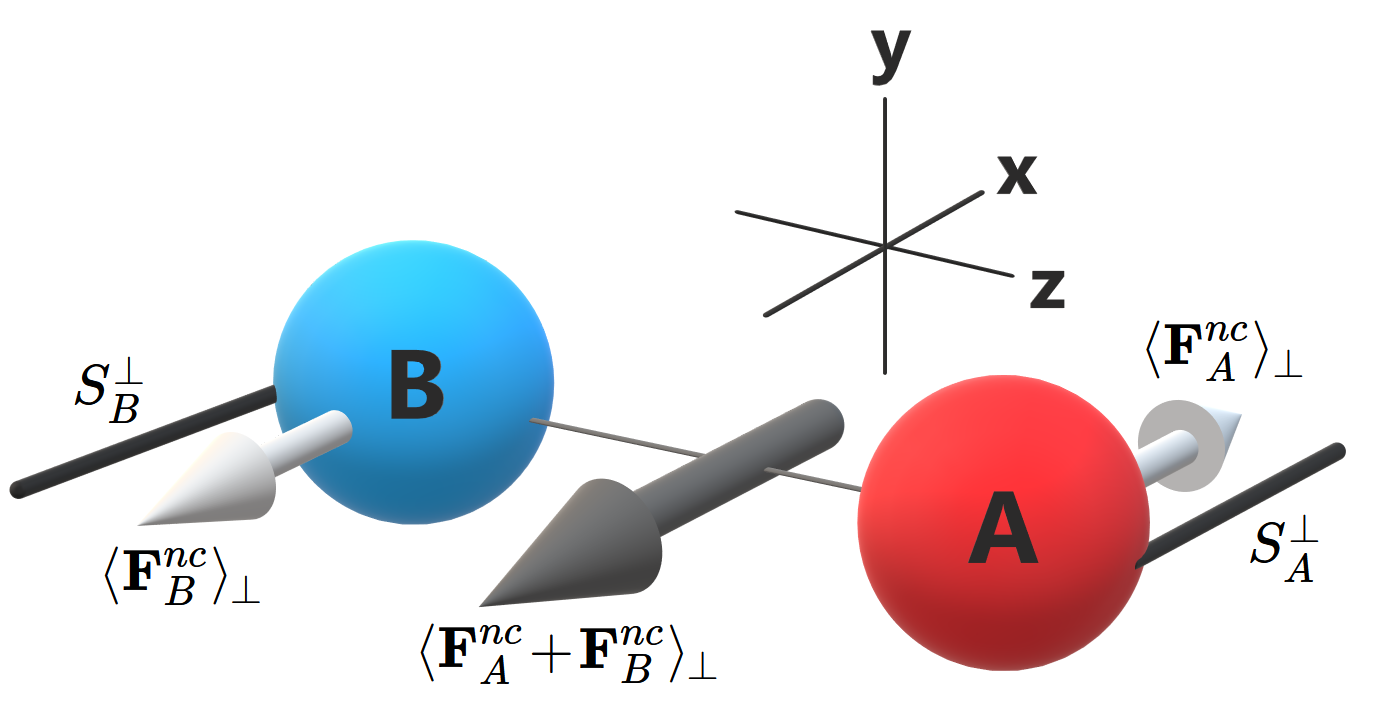}
\caption{Pictorial representation of the action of the orthogonal components of the nonconservative forces upon a binary atomic 
system which causes the displacements of the atoms in a direction orthogonal to the interatomic axis, $S_{A,B}^{\perp}$.}\label{artistforces}
\end{figure}
The strength of the nonconservative forces, in the perturbative regime, are of an order $\Gamma_{0}/\omega_{0}$ weaker than 
the vdW forces \cite{Julio_PRA1}. Hence, the components along the interatomic axis are hardly distinguishable experimentally. On the contrary, their orthogonal 
components, absent in the vdW forces, might be observed. The orthogonal components of the reciprocal forces generate a torque around the 
center of mass, while the net force of 
Eq.(\ref{Fnc_net}) displaces the center of mass as illustrated in Fig.\ref{artistforces}. 

\begin{figure}[H]
\includegraphics[height=6.cm,width=8.4cm,clip]{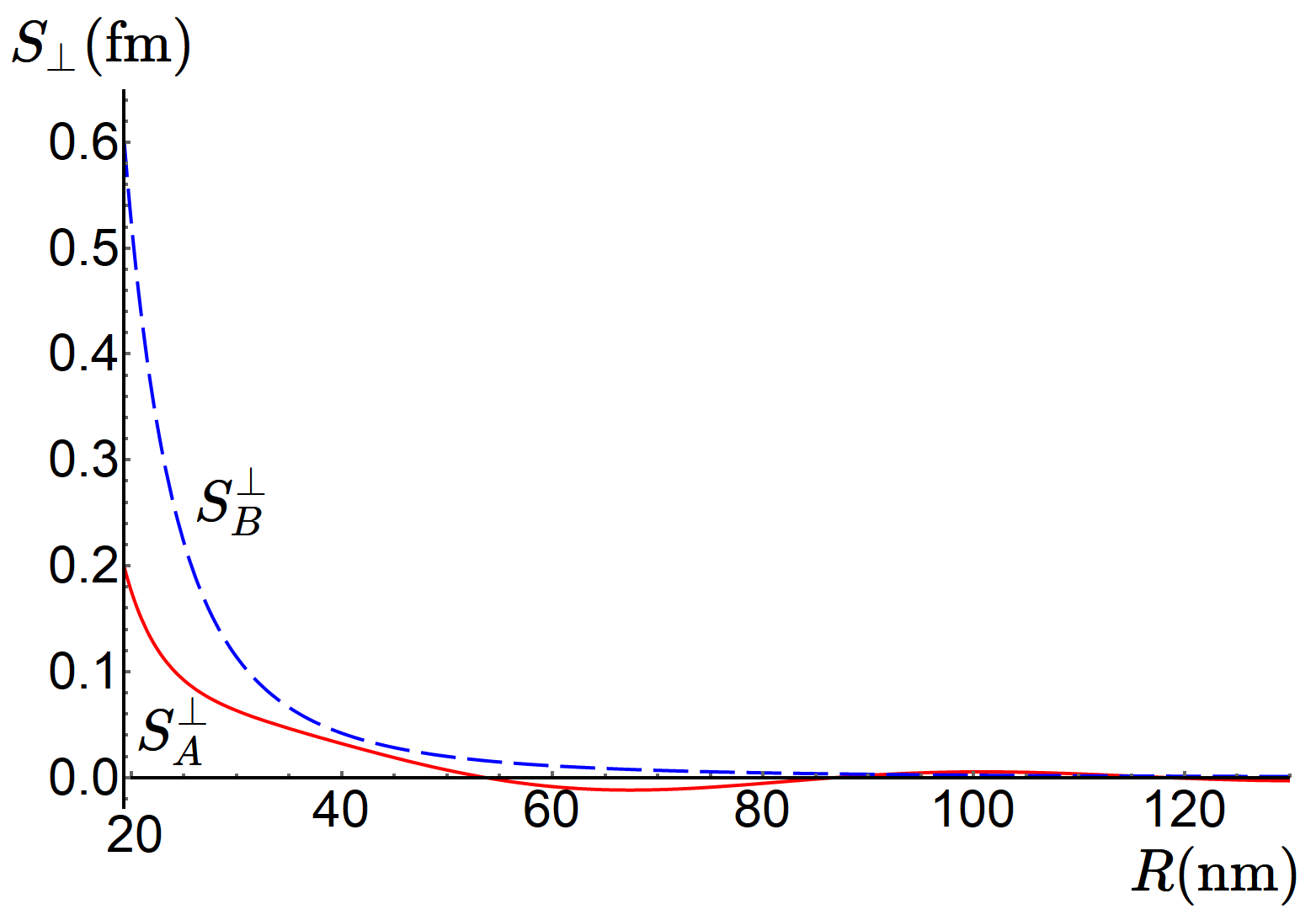}
\caption{Graphical representation of the perpendicular displacements along the $-x$-axis caused by the nonconservative dipole forces on a binary system of Hydrogen atoms, 
with one of them, $A$, initially excited to the state $(2p_{x}+2p_{z})/\sqrt{2}$, as a function of the interatomic distance $R$ along the $z$-axis. The solid line 
in red represents $S_{A}^{\perp}$, while the dashed line in blue is for  $S_{B}^{\perp}$, according to Eq.(\ref{displ}).}\label{Hidrogen}
\end{figure}

We finalize with the estimate of the displacement caused on an excited binary system of Hydrogen atoms by the 
orthogonal components of the nonconservative forces. Considering the atom $A$ initially excited to the state 
$|A_{+}\rangle=(2p_{x}+2p_{z})/\sqrt{2}$, and that the atoms are placed a distance $R$ apart along the $\hat{\mathbf{z}}$ axis --see Fig.\ref{artistforces}, 
the displacement $\mathbf{S}_{A,B}^{\perp}$ of each atom along the $\hat{\mathbf{x}}$ axis in a lifetime $\sim1.6~$ns as a function of the interatomic distance is 
\begin{align}
\mathbf{S}_{A}^{\perp}&\simeq0.15~\textrm{fm}\frac{(2v^2-1)\cos{2v}-(2v-v^3)\sin{2v}}{v^5}\hat{\mathbf{x}},\nonumber\\
\mathbf{S}_{B}^{\perp}&\simeq-0.3~\textrm{fm}\frac{1+v^{2}}{v^5}\hat{\mathbf{x}},\: v\equiv2\pi R/\lambda_{0},\label{displ}
\end{align}
for $\lambda_{0}\simeq121.6~$nm. In order for our computation to remain perturbative, $v\gtrsim1$, meaning that the maximum values of the perpendicular displacements 
are of the order of 1~fm in the perturbative regime --see Fig.\ref{Hidrogen}. Note also that for $R\lesssim50$ nm both atoms move in the same direction, 
meaning that the nonreciprocal components of the forces dominate there.
\section{Conclusions and outlook} 
We have computed the nonconservative dipole forces between two two-level identical atoms, one of which is initially excited. We have found that these forces are 
of an order $\Gamma_{0}/\omega_{0}$ weaker than the vdW forces. Nonetheless, they posses components orthogonal to the interatomic axis that might 
be experimentally accessible.  Our perturbative computation on a binary system of Hydrogen atoms shows that the orthogonal displacement of the atoms is of the order 
of a fermi in the middle-far field regime. A nonperturbative computation would be necessary in the near field, where the displacement is expected to be greater. 
In this respect, in order to facilitate its observation, Rydberg atoms present themselves as good candidates 
\cite{Haroche1,Reinhard2007,Forster,Weber,Scheel,somepapers,Beguin}. 

As with the vdW forces, the leading terms of the nonconservative forces are fully resonant, and contain nonreciprocal 
components that generate a net displacement of the two-atom system. However, in contrast to the net vdW force, the net nonconservative force 
is not related to directionality of spontaneous emission, but equals itself the time variation of 
the longitudinal EM momentum \cite{comment}.  

\acknowledgments
Financial support is acknowledged from grants MTM2014-57129-C2-1-P (MINECO) and VA137G18, BU229P18 (JCyL), as well as from the project QCAYLE (NextGenerationEU funds).

\appendix

\begin{widetext}

\section{Complete formulas of the nonconservative forces}\label{appendix}

In this Appendix we compile the complete expressions of the  nonconservative electric dipole forces between two atoms, $A$ and $B$, one which, say $A$, 
is  initially excited. In the first place, we address the calculation for dissimilar atoms. Later, we consider the identical atoms limit which leads to the 
equations included in the main text.

\subsection{Nonconservative forces on dissimilar atoms}

In terms of the nomenclature used in the main text, the atoms $A$ and $B$ are said dissimilar if $|\Delta_{AB}|\gg\Gamma_{A,B}$. Following the time-dependent 
perturbative approach outlined in the Sec.\ref{fundam}, we compute the contribution of each of the diagrams in Fig.1.  Since we are ultimately interested in the identical atoms 
limit, we restrict ourselves to the sudden excitation approximation used in Ref.\cite{Julio_PRA1}. 

In the first place, we illustrate the calculation with the detailed reading of the contribution of diagram (a) to $\langle\mathbf{F}_{A}^{nc}\rangle_{T}$. It reads 
\begin{align}\label{laeq}
&\frac{1}{\hbar^{3}}\frac{\partial}{\partial T}\int_{0}^{\infty}\frac{\mathcal{V}k^{2}\textrm{d}k}{(2\pi)^{3}}\int_{0}^{\infty}\frac{\mathcal{V}k^{'2}\textrm{d}k'}
{(2\pi)^{3}}\int_{0}^{4\pi}\textrm{d}\Theta\int_{0}^{4\pi}\textrm{d}\Theta'\Bigl\{\Bigl[i\langle A_{+},B_{-},0_{\gamma}|e^{i \Omega_{a}^{*}T}
| A_{+},B_{-},0_{\gamma}\rangle\int_{0}^{T}\textrm{d}t\int_{0}^{t}\textrm{d}t'\int_{0}^{t'}\textrm{d}t''\nonumber\\
&\times\sum_{b}\langle  A_{+},B_{-},0_{\gamma}|\mathbf{d}_{A}\times\mathbf{B}^{(-)}_{\mathbf{k}}(\mathbf{R}_{A})|A_{-},B_{-},\gamma_{\mathbf{k}}
\rangle e^{-i\omega(T-t)}\langle A_{-},B_{-},\gamma_{\mathbf{k}}|\mathbf{d}_{B}\cdot\mathbf{E}_{\mathbf{k}}^{(+)}(\mathbf{R}_{B})|A_{-},b,0_{\gamma}\rangle\nonumber\\
&\times e^{-i \Omega_{b}(t-t')}\langle A_{-},b,0_{\gamma}|\mathbf{d}_{B}\cdot\mathbf{E}_{\mathbf{k}'}^{(-)}(\mathbf{R}_{B})|A_{-},B_{-},\gamma_{\mathbf{k}'}\rangle 
e^{-i\omega'(t'-t'')}\langle A_{-},B_{-},\gamma_{\mathbf{k}'}|\mathbf{d}_{A}\cdot\mathbf{E}_{\mathbf{k}'}^{(+)}(\mathbf{R}_{A})| A_{+},B_{-},0_{\gamma}\rangle e^{-i \Omega_{a} t''}\Bigr]\nonumber\\
&+[k\leftrightarrow k']^{\dagger}\Bigr\},\nonumber\qquad\qquad\qquad\qquad\qquad\qquad\qquad\qquad\qquad\qquad\qquad\qquad\qquad\qquad\qquad\qquad\qquad\qquad
\qquad\qquad\qquad(\textrm{A.1})
\end{align}  
where  $|A_{+},B_{-},0_{\gamma}\rangle$ is the initial two-atom-EM-vacuum state, with atom $A$ excited at time 0, $|\gamma_{\mathbf{k}}\rangle$ is a one-photon state 
of momentum $\mathbf{k}$ and frequency $\omega=ck$, $\mathcal{V}$ is the volume of quantization to be taken eventually to infinity,  $\Theta$ and $\Theta^{'}$ are the solid angle variables, and the complex time-exponentials are the result of the application of the free time-evolution operator 
$\mathbb{U}_{0}(t)=e^{-i\hbar^{-1}H_{0}t}$ between the interaction 
vertices $W_{A,B}$, with $ \Omega_{a}=\omega_{A}-i\Gamma_{A}/2$ and $ \Omega_{b}=\omega_{B}-i\Gamma_{b}/2$, where the dissipative imaginary terms account for 
radiative emission in the Weisskopf-Wigner approximation. Integrating in time and solid angles the 
expression of Eq.(A.1), one obtains  
\begin{align}
&\frac{-c}{ \hbar\pi^{2}\epsilon_{0}^{2}}\frac{\partial}{\partial T}\sum_{b}\textrm{Re}\int_{0}^{\infty}ikdk\:\boldsymbol{\mu}_{A}\times\boldsymbol{\nabla}_{A}
\times\textrm{Im}\mathbb{G}(kR)\cdot\boldsymbol{\mu}_{b}\int_{0}^{\infty}dk'k^{'2}\boldsymbol{\mu}_{B}\cdot\textrm{Im}\mathbb{G}(k'R)\cdot\boldsymbol{\mu}_{A}
\:e^{i \Omega_{a}^{*}T}\Bigl[ \frac{e^{-i  \Omega_{a} T}-e^{-i \omega T}}{(\omega'- \Omega_{a})( \Omega_{b}- \Omega_{a})(\omega -  \Omega_{a})}\nonumber\\
&-\frac{e^{-i  \Omega_{b} T}-e^{-i \omega T}}{(\omega'- \Omega_{a})( \Omega_{b}- \Omega_{a})(\omega -  \Omega_{b})}+\frac{e^{-i \omega' T}-e^{-i \omega T}}
{(\omega'- \Omega_{a})(\omega'- \Omega_{b})(\omega - \omega')}-\frac{e^{-i  \Omega_{b} T}-e^{-i \omega T}}{(\omega'- \Omega_{a})(\omega'- \Omega_{b})
(\omega -\Omega_{b})} \Bigr],\qquad\qquad\qquad\:\:\quad\textrm{(A.2)}\nonumber
\end{align}
where the notation is that of the main text. 

Operating in a similar manner with the rest of the diagrams of 
Fig.1 for $\langle\mathbf{F}_{A}^{nc}\rangle_{T}$ and their analogous for $\langle\mathbf{F}_{B}^{nc}\rangle_{T}$,  
and upon integration in $k$ and $k'$ in the complex plane, we arrive at 
\begin{equation}
\begin{split}
&\langle\mathbf{F}_A^{nc}\rangle_{T}=\sum_{b}\Bigl\{\\&\frac{2\omega_A^3 \Gamma_A e^{-\Gamma_A T}}{c^4 \epsilon_0^2\hbar\Delta_{AB}}\Bigl[\boldsymbol{\mu}_{A}\times \boldsymbol{\nabla} \times\textrm{Re}\mathbb{G}(k_A R)\cdot\boldsymbol{\mu}_{b}
\boldsymbol{\mu}_{b}\cdot\textrm{Im}\mathbb{G}(k_A R)\cdot\boldsymbol{\mu}_{A}+\boldsymbol{\mu}_{A}\times \boldsymbol{\nabla} \times \textrm{Im}
\mathbb{G}(k_A R)\cdot\boldsymbol{\mu}_{b}\boldsymbol{\mu}_{b}\cdot\textrm{Re}\mathbb{G}(k_A R)\cdot\boldsymbol{\mu}_{A}\Bigr]\\
&+\frac{\omega_B^3\left[2 \Delta_{AB}-(\Gamma_A+\Gamma_{b})\right] e^{-(\Gamma_A+\Gamma_{b})T/2}}{c^4 \epsilon_0^2\hbar\Delta_{AB}} \\& \times\Bigl[\boldsymbol{\mu}_{A}\times \boldsymbol{\nabla} \times\textrm{Re}\mathbb{G}(k_B R)\cdot\boldsymbol{\mu}_{b}
\boldsymbol{\mu}_{b}\cdot\textrm{Im}\mathbb{G}(k_B R)\cdot\boldsymbol{\mu}_{A}+\boldsymbol{\mu}_{A}\times \boldsymbol{\nabla} \times \textrm{Im}
\mathbb{G}(k_B R)\cdot\boldsymbol{\mu}_{b}\boldsymbol{\mu}_{b}\cdot\textrm{Re}\mathbb{G}(k_B R)\cdot\boldsymbol{\mu}_{A}\Bigr]\cos(\Delta_{AB}T)\nonumber\\
&-\frac{\omega_B^3\left[2\Delta_{AB}+(\Gamma_A+\Gamma_{b})\right] e^{-(\Gamma_A+\Gamma_{b})T/2}}{c^4 \epsilon_0^2\hbar\Delta_{AB}}\\& \times\Bigl[\boldsymbol{\mu}_{A}\times \boldsymbol{\nabla} \times\textrm{Re}\mathbb{G}(k_B R)\cdot\boldsymbol{\mu}_{b}
\boldsymbol{\mu}_{b}\cdot\textrm{Re}\mathbb{G}(k_B R)\cdot\boldsymbol{\mu}_{A}-\boldsymbol{\mu}_{A}\times \boldsymbol{\nabla} \times \textrm{Im}
\mathbb{G}(k_B R)\cdot\boldsymbol{\mu}_{b}\boldsymbol{\mu}_{b}\cdot\textrm{Im}\mathbb{G}(k_B R)\cdot\boldsymbol{\mu}_{A}\Bigr]\sin(\Delta_{AB}T)\\
&-\frac{2  \omega_A^3 \Gamma_A e^{-\Gamma_A T}}{c^4 \epsilon_0^2 \hbar(\omega_A+\omega_B)}\Bigl[\boldsymbol{\mu}_{A}\times \boldsymbol{\nabla} \times\textrm{Re}\mathbb{G}(k_A R)\cdot\boldsymbol{\mu}_{b}
\boldsymbol{\mu}_{b}\cdot\textrm{Im}\mathbb{G}(k_A R)\cdot\boldsymbol{\mu}_{A}+\boldsymbol{\mu}_{A}\times \boldsymbol{\nabla} \times \textrm{Im}
\mathbb{G}(k_A R)\cdot\boldsymbol{\mu}_{b}\boldsymbol{\mu}_{b}\cdot\textrm{Re}\mathbb{G}(k_A R)\cdot\boldsymbol{\mu}_{A}\Bigr]\\
&+\frac{\omega_B(\Gamma_A+\Gamma_b)  e^{-(\Gamma_A+\Gamma_b) T/2} }{c^3 \epsilon_0^2 \hbar}\Bigl[\boldsymbol{\mu}_{A}\times \boldsymbol{\nabla} \times\textrm{Im}\mathbb{G}(k_B R)\cdot\boldsymbol{\mu}_{b} \cos(\Delta_{AB}T)+\boldsymbol{\mu}_{A}\times \boldsymbol{\nabla} \times\textrm{Re}\mathbb{G}(k_B R)\cdot\boldsymbol{\mu}_{b}\sin(\Delta_{AB}T)\Bigr]\\
&\times \int_{0}^{\infty}\frac{dq}{\pi} \frac{(q^2-k_A k_B)q^2\boldsymbol{\mu}_{A}\cdot\mathbb{G}(iq R)\cdot\boldsymbol{\mu}_{b}}{(q^2+k_A^2)(q^2+k_B^2)}\\
&+\frac{2 \omega_B \Delta_{AB}  e^{-(\Gamma_A+\Gamma_b) T/2} }{c^3 \epsilon_0^2 \hbar}\Bigl[\boldsymbol{\mu}_{A}\times \boldsymbol{\nabla} \times\textrm{Im}\mathbb{G}(k_B R)\cdot\boldsymbol{\mu}_{b} \sin(\Delta_{AB}T)-\boldsymbol{\mu}_{A}\times \boldsymbol{\nabla} \times\textrm{Re}\mathbb{G}(k_B R)\cdot\boldsymbol{\mu}_{b}\cos(\Delta_{AB}T)\Bigr]\\
&\times \int_{0}^{\infty}\frac{dq}{\pi} \frac{(q^2-k_A k_B)q^2\boldsymbol{\mu}_{A}\cdot\mathbb{G}(iq R)\cdot\boldsymbol{\mu}_{b}}{(q^2+k_A^2)(q^2+k_B^2)}\Bigr\},
\qquad\qquad\qquad\qquad\qquad\qquad\qquad\qquad\qquad\qquad\qquad\qquad\qquad\qquad\:\:\quad\textrm{(A.3)}\nonumber
\end{split}
\end{equation}
\begin{equation}
\begin{split}
&\langle\mathbf{F}_B^{nc}\rangle_{T}=\sum_{b}\Bigl\{\\&-\frac{2\omega_A^3\Gamma_A e^{-\Gamma_A T}}{c^4 \epsilon_0^2\hbar\Delta_{AB}}\Bigl[\boldsymbol{\mu}_{b}\times\boldsymbol{\nabla} \times\textrm{Re}\mathbb{G}(k_A R)\cdot\boldsymbol{\mu}_{A}
\boldsymbol{\mu}_{b}\cdot\textrm{Im}\mathbb{G}(k_A R)\cdot\boldsymbol{\mu}_{A}-\boldsymbol{\mu}_{b}\times \boldsymbol{\nabla} \times \textrm{Im}
\mathbb{G}(k_A R)\cdot\boldsymbol{\mu}_{A}\boldsymbol{\mu}_{b}\cdot\textrm{Re}\mathbb{G}(k_A R)\cdot\boldsymbol{\mu}_{A}\Bigr]\\
&-\frac{\omega_A\omega_B^2 \left[2\Delta_{AB}-(\Gamma_A+\Gamma_{b})\right] e^{-(\Gamma_A+\Gamma_{b})T/2}}{c^4 \epsilon_0^2\hbar\Delta_{AB}} \\& \times\Bigl[\boldsymbol{\mu}_{b}\times \boldsymbol{\nabla} \times\textrm{Re}\mathbb{G}(k_A R)\cdot\boldsymbol{\mu}_{A}
\boldsymbol{\mu}_{b}\cdot\textrm{Im}\mathbb{G}(k_B R)\cdot\boldsymbol{\mu}_{A}-\boldsymbol{\mu}_{b}\times \boldsymbol{\nabla} \times \textrm{Im}
\mathbb{G}(k_A R)\cdot\boldsymbol{\mu}_{A}\boldsymbol{\mu}_{b}\cdot\textrm{Re}\mathbb{G}(k_B R)\cdot\boldsymbol{\mu}_{A}\Bigr]\cos(\Delta_{AB}T)\\
&+\frac{\omega_A\omega_B^2\left[2\Delta_{AB}+(\Gamma_A+\Gamma_{b})\right] e^{-(\Gamma_A+\Gamma_{b})T/2}}{c^4 \epsilon_0^2\hbar\Delta_{AB}}\\&  \times\Bigl[\boldsymbol{\mu}_{b}\times \boldsymbol{\nabla} \times\textrm{Re}\mathbb{G}(k_A R)\cdot\boldsymbol{\mu}_{A}
\boldsymbol{\mu}_{b}\cdot\textrm{Re}\mathbb{G}(k_B R)\cdot\boldsymbol{\mu}_{A}+\boldsymbol{\mu}_{b}\times \boldsymbol{\nabla} \times \textrm{Im}
\mathbb{G}(k_A R)\cdot\boldsymbol{\mu}_{A}\boldsymbol{\mu}_{b}\cdot\textrm{Im}\mathbb{G}(k_B R)\cdot\boldsymbol{\mu}_{A}\Bigr]\sin(\Delta_{AB}T)\\
&+\frac{2  \omega_A^3 \Gamma_A e^{-\Gamma_A T}}{c^4 \epsilon_0^2 \hbar(\omega_A+\omega_B)}\Bigl[\boldsymbol{\mu}_{b}\times \boldsymbol{\nabla} \times\textrm{Re}\mathbb{G}(k_A R)\cdot\boldsymbol{\mu}_{A}
\boldsymbol{\mu}_{b}\cdot\textrm{Im}\mathbb{G}(k_A R)\cdot\boldsymbol{\mu}_{A}-\boldsymbol{\mu}_{b}\times \boldsymbol{\nabla} \times \textrm{Im}
\mathbb{G}(k_A R)\cdot\boldsymbol{\mu}_{A}\boldsymbol{\mu}_{b}\cdot\textrm{Re}\mathbb{G}(k_A R)\cdot\boldsymbol{\mu}_{A}\Bigr]\\
&+\frac{\omega_A (\Gamma_A+\Gamma_b) e^{-(\Gamma_A+\Gamma_b) T/2} }{c^3 \epsilon_0^2 \hbar}\Bigl[\boldsymbol{\mu}_{b}\times \boldsymbol{\nabla} \times\textrm{Im}\mathbb{G}(k_A R)\cdot\boldsymbol{\mu}_{A} \cos(\Delta_{AB}T)-\boldsymbol{\mu}_{b}\times \boldsymbol{\nabla} \times\textrm{Re}\mathbb{G}(k_A R)\cdot\boldsymbol{\mu}_{A}\sin(\Delta_{AB}T)\Bigr]\\
&\times \int_{0}^{\infty}\frac{dq}{\pi} \frac{(q^2-k_A k_B)q^2\boldsymbol{\mu}_{A}\cdot\mathbb{G}(iq R)\cdot\boldsymbol{\mu}_{b}}{(q^2+k_A^2)(q^2+k_B^2)}\\
&+\frac{2 \omega_A \Delta_{AB}  e^{-(\Gamma_A+\Gamma_b) T/2} }{c^3 \epsilon_0^2 \hbar}\Bigl[\boldsymbol{\mu}_{b}\times \boldsymbol{\nabla} \times\textrm{Im}\mathbb{G}(k_A R)\cdot\boldsymbol{\mu}_{A} \sin(\Delta_{AB}T)+
\boldsymbol{\mu}_{b}\times \boldsymbol{\nabla} \times\textrm{Re}\mathbb{G}(k_A R)\cdot\boldsymbol{\mu}_{A} \cos(\Delta_{AB}T)\Bigr]\\
&\times \int_{0}^{\infty}\frac{dq}{\pi} \frac{(q^2-k_A k_B)q^2\boldsymbol{\mu}_{A}\cdot\mathbb{G}(iq R)\cdot\boldsymbol{\mu}_{b}}{(q^2+k_A^2)(q^2+k_B^2)}\Bigr\}.
\qquad\qquad\qquad\qquad\qquad\qquad\qquad\qquad\qquad\qquad\qquad\qquad\qquad\qquad\:\:\quad\textrm{(A.4)}\nonumber
\end{split}
\end{equation}
The origin of each term is as follows. The time oscillating terms of frequency 
$\Delta_{AB}$ arise from diagram (a). They contain the dominant contribution, together with terms of an order $\Gamma_{A,b}/\Delta_{AB}$ smaller. In addition, 
the quasi-stationary terms of an order $\Gamma_{A,b}/\omega_{A,B}$ less come from diagram (g), and semi-resonant terms arise from diagrams (c) and (d). 
Fast oscillating spurious terms of frequency $(\omega_A+\omega_B)$ which arise from diagrams (k) and (l) are neglected. Note that, in contrast to the conservative 
vdW forces, there are no fully off-resonant components. 

\subsection{Nonconservative forces on identical atoms}
Next, considering the identical atoms limit upon the expressions in Eqs.(A.3) and (A.4), $\Gamma_{A}\rightarrow\Gamma_{b}\equiv\Gamma_{0}$, $\omega_{A}\rightarrow\omega_{B}\equiv\omega_{0}$, 
$\Delta_{AB}/\Gamma_{0}\rightarrow0$, the nonconservative forces on each atom read,
\begin{equation}
\begin{split}
&\langle \mathbf{F}_A^{nc}\rangle_{T}=\sum_{b}\Bigl\{\\&\frac{2\Gamma_0 e^{-\Gamma_0 T}}{c^4 \epsilon_0^2\hbar} \frac{\partial}{\partial \omega} 
\Bigl[\omega^3[\boldsymbol{\mu}_{A}\times \boldsymbol{\nabla} \times\textrm{Re}\mathbb{G}(k R)\cdot\boldsymbol{\mu}_{b}
\boldsymbol{\mu}_{b}\cdot\textrm{Im}\mathbb{G}(k R)\cdot\boldsymbol{\mu}_{A}+\boldsymbol{\mu}_{A}\times \boldsymbol{\nabla} \times \textrm{Im}
\mathbb{G}(k R)\cdot\boldsymbol{\mu}_{b}\boldsymbol{\mu}_{b}\cdot\textrm{Re}\mathbb{G}(k R)\cdot\boldsymbol{\mu}_{A}]\Bigr]_{\omega=\omega_0}\\
&+\frac{2\omega_0^3(1-\Gamma_0T) e^{-\Gamma_0T}}{c^4 \epsilon_0^2\hbar}\Bigl[\boldsymbol{\mu}_{A}\times \boldsymbol{\nabla} \times\textrm{Re}\mathbb{G}(k_0 R)\cdot\boldsymbol{\mu}_{b}
\boldsymbol{\mu}_{b}\cdot\textrm{Re}\mathbb{G}(k_0 R)\cdot\boldsymbol{\mu}_{A}-\boldsymbol{\mu}_{A}\times \boldsymbol{\nabla} \times \textrm{Im}
\mathbb{G}(k_0 R)\cdot\boldsymbol{\mu}_{b}\boldsymbol{\mu}_{b}\cdot\textrm{Im}\mathbb{G}(k_0 R)\cdot\boldsymbol{\mu}_{A}\Bigr]\\
&-\frac{2  \omega_0^2 \Gamma_0 e^{-\Gamma_0 T}}{c^4 \epsilon_0^2 \hbar}\Bigl[\boldsymbol{\mu}_{A}\times \boldsymbol{\nabla} \times\textrm{Re}\mathbb{G}(k_0 R)\cdot\boldsymbol{\mu}_{b}
\boldsymbol{\mu}_{b}\cdot\textrm{Im}\mathbb{G}(k_0 R)\cdot\boldsymbol{\mu}_{A}+\boldsymbol{\mu}_{A}\times \boldsymbol{\nabla} \times \textrm{Im}
\mathbb{G}(k_0 R)\cdot\boldsymbol{\mu}_{b}\boldsymbol{\mu}_{b}\cdot\textrm{Re}\mathbb{G}(k_0 R)\cdot\boldsymbol{\mu}_{A}\Bigr]\\
&+\frac{2 \omega_0 \Gamma_0  e^{-\Gamma_0 T} }{c^3 \epsilon_0^2 \hbar}\boldsymbol{\mu}_{A}\times \boldsymbol{\nabla}
\times\textrm{Im}\mathbb{G}(k_0 R)\cdot\boldsymbol{\mu}_{b} \int_{0}^{\infty}\frac{dq}{\pi} \frac{(q^2-k_0^2)q^2\boldsymbol{\mu}_{A}\cdot
\mathbb{G}(iq R)\cdot\boldsymbol{\mu}_{b}}{(q^2+k_0^2)^2}\Bigr\},
\quad\qquad\qquad\qquad\qquad\qquad\qquad\:\:\textrm{(A.5)}\nonumber
\end{split}
\end{equation}	
\begin{equation}
\begin{split}
&\langle  \mathbf{F}_B^{nc}\rangle_{T}=\sum_{b}\Bigl\{-\frac{2\Gamma_0 \omega_0 e^{-\Gamma_0 T}}{c^4 \epsilon_0^2\hbar}\Bigl[\boldsymbol{\mu}_{b}\times
\boldsymbol{\nabla}\times\textrm{Re}\mathbb{G}(k R)\cdot\boldsymbol{\mu}_{A} \frac{\partial}{\partial \omega} \left[\omega^2
\boldsymbol{\mu}_{b}\cdot\textrm{Im}\mathbb{G}(k R)\cdot\boldsymbol{\mu}_{A}\right]_{\omega=\omega_0}\\
&+\boldsymbol{\mu}_{b}\times
\boldsymbol{\nabla} \times \textrm{Im}\mathbb{G}(k R)\cdot\boldsymbol{\mu}_{A}\frac{\partial}{\partial \omega}
\left[ \omega^2\boldsymbol{\mu}_{b}\cdot\textrm{Re}\mathbb{G}(k R)\cdot\boldsymbol{\mu}_{A} \right]_{\omega=\omega_0}\Bigr]\\
&-\frac{2\omega_0^3 (1-\Gamma_0T) e^{-\Gamma_0T}}{c^4 \epsilon_0^2\hbar}\Bigl[\boldsymbol{\mu}_{b}\times \boldsymbol{\nabla}
\times\textrm{Re}\mathbb{G}(k_0 R)\cdot\boldsymbol{\mu}_{A}
\boldsymbol{\mu}_{b}\cdot\textrm{Re}\mathbb{G}(k_0 R)\cdot\boldsymbol{\mu}_{A}+\boldsymbol{\mu}_{b}\times \boldsymbol{\nabla} \times \textrm{Im}
\mathbb{G}(k_0 R)\cdot\boldsymbol{\mu}_{A}\boldsymbol{\mu}_{b}\cdot\textrm{Im}\mathbb{G}(k_0 R)\cdot\boldsymbol{\mu}_{A}\Bigr]\\
&+\frac{2\omega_0^2 \Gamma_0 e^{-\Gamma_0 T}}{c^4 \epsilon_0^2 \hbar}\Bigl[\boldsymbol{\mu}_{b}\times \boldsymbol{\nabla} \times\textrm{Re}\mathbb{G}(k_0 R)\cdot\boldsymbol{\mu}_{A}
\boldsymbol{\mu}_{b}\cdot\textrm{Im}\mathbb{G}(k_0 R)\cdot\boldsymbol{\mu}_{A}-\boldsymbol{\mu}_{b}\times \boldsymbol{\nabla} \times \textrm{Im}
\mathbb{G}(k_0 R)\cdot\boldsymbol{\mu}_{A}\boldsymbol{\mu}_{b}\cdot\textrm{Re}\mathbb{G}(k_0 R)\cdot\boldsymbol{\mu}_{A}\Bigr]\\
&+\frac{2 \omega_0 \Gamma_0  e^{-\Gamma_0 T} }{c^3 \epsilon_0^2 \hbar}\boldsymbol{\mu}_{b}\times \boldsymbol{\nabla}
\times\textrm{Im}\mathbb{G}(k_0 R)\cdot\boldsymbol{\mu}_{A} \int_{0}^{\infty}\frac{dq}{\pi} \frac{(q^2-k_0^2)q^2\boldsymbol{\mu}_{A}\cdot
\mathbb{G}(iq R)\cdot\boldsymbol{\mu}_{b}}{(q^2+k_0^2)^2}\Bigr\}.
\quad\qquad\qquad\qquad\qquad\qquad\qquad\quad\textrm{(A.6)}\nonumber
\end{split}
\end{equation}
The dominant terms of these equations are those in Eq.(6) of the main text, while the rest are higher order corrections of strengths $\Gamma_{0}/\omega_{0}$ and 
$R/cT$ times smaller. Finally, the following identities have been used in the Letter in order to differentiate the axial components from the orthogonal components 
in the nonconservative forces,
\begin{equation}
 ik \,\ \mathcal{G}(kR)\mathcal{E} \cdot \hat{\mathbf{R}}= \boldsymbol{\nabla} \times \mathbb{G}(kR),\quad
\boldsymbol{\mu}_{A,b} \times\boldsymbol{\nabla} \times \mathbb{G}(kR) \cdot\boldsymbol{\mu}_{b,A} = i k (\mu_{A,b}^{\parallel}\boldsymbol{\mu}_{b,A}^{\perp}-\mu_A^{\perp}\mu_b^{\perp}
\hat{\mathbf{R}}) \mathcal{G}(kR),\nonumber
\end{equation}
where $\mathbb{G}(kr)$ is the dyadic Green's function of the electric field induced at $\mathbf{r}$ by an electric dipole of frequency $ck$ at the origin; 
$\mathcal{G}(kr)\mathcal{E}\cdot\hat{\mathbf{r}}$ is the 
dyadic Green's tensor of the magnetic field induced at $\mathbf{r}$ by an electric dipole of frequency $ck$ at the origin, with $\mathcal{E}$ being the 
three dimensional Levi-Civita tensor. Their imaginary parts correspond to the vacuum expectation values of the quadratic fluctuations of the electromagnetic field 
which appear in Eq.(A.2), 
\begin{equation}
\begin{split}
&\int d\Theta_{\mathbf{k}}\langle0_{\gamma}|\mathbf{E}^{(-)}_{\mathbf{k}}(\mathbf{r})\:\mathbf{E}^{(+)}_{\mathbf{k}}(\mathbf{0})|0_{\gamma}\rangle=
-\frac{8 \pi^2 \hbar c}{\epsilon_0} \,\ \textrm{Im}\:\mathbb{G}(kr),
\\ &\int d\Theta_{\mathbf{k}}\langle0_{\gamma}|\mathbf{B}^{(-)}_{\mathbf{k}}(\mathbf{r})\:\mathbf{E}^{(+)}_{\mathbf{k}}(\mathbf{0})|0_{\gamma}\rangle=
-\frac{8 \pi^2 i\hbar}{\epsilon_0 k} \,\ \boldsymbol{\nabla}_{\mathbf{R}}\times\textrm{Im}\:\mathbb{G}\nonumber(kr).
\nonumber
\end{split}
\end{equation}
\end{widetext}


\begin{thebibliography}{105}
\bibitem{comment} In this respect, forces related to the time variation of the longitudinal EM momentum have been 
also estimated in magnetochiral systems \cite{KawkavanTiggelenRikken,myPRLchiral,JPCM}.
\end{thebibliography}
\end{document}